\documentclass[twocolumn,secnumarabic,amssymb, nobibnotes, aps, prd, superscriptaddress]{revtex4-1}

\usepackage{graphicx}
\usepackage{color}

\DeclareTextSymbol{\degres}{OT1}{23}

\begin{document}

\title{Influence of grafting on the glass transition temperature of PS thin films}

\author{Marceau H\'enot}
\affiliation{Laboratoire de Physique des Solides, CNRS, Univ. Paris-Sud, Universit\'e
Paris-Saclay, 91405 Orsay Cedex, France}
\author{Alexis Chennevi\`ere}
\affiliation{Laboratoire de Physique des Solides, CNRS, Univ. Paris-Sud, Universit\'e
Paris-Saclay, 91405 Orsay Cedex, France}
\author{Eric Drockenmuller}
\affiliation{Univ Lyon, Universit\'e Lyon 1, CNRS, Ing\'enierie des Mat\'eriaux Polym\`eres, UMR 5223, F-69003, LYON, France}
\author{Kenneth Shull}
\affiliation{Department of Materials Science and Engineering, Northwestern University, 2220 Campus Drive, Evanston, Illinois 60208, United States}
\author{Liliane L\'eger}
\affiliation{Laboratoire de Physique des Solides, CNRS, Univ. Paris-Sud, Universit\'e
Paris-Saclay, 91405 Orsay Cedex, France}
\author{Fr\'ed\'eric Restagno}
\email[Corresponding author : ]{frederic.restagno@u-psud.fr}
\affiliation{Laboratoire de Physique des Solides, CNRS, Univ. Paris-Sud, Universit\'e
Paris-Saclay, 91405 Orsay Cedex, France}
\date{\today}

\begin{abstract}
We present an investigation of the effect of the interaction between a thin polystyrene film and its supporting substrate on its glass transition temperature ($T_{\mathrm{g}}$).  We modulate this interaction by depositing the film on end-tethered polystyrene grafted layers of controlled molecular parameters. By comparing $T_{\mathrm{g}}$ measurements versus film thickness for films deposited on different grafted layers and films deposited directly on a silicon substrate, we can conclude that there is no important effect of the film-subtrate interaction. Our interpretation of these results is that local orientation and dynamic effects substantial enough to influence $T_{\mathrm{g}}$ cannot readily be obtained by grafting prepolymerized chains to a surface, due to intrinsic limitation of the surface grafting density.
\end{abstract}
\maketitle

\section{Introduction}
The properties of polymers in nanometric films are of major importance in various high performance materials such as nanocomposites, multilayer materials or lubricants. Those materials are often used close to their glass transition temperature ($T_{\mathrm{g}}$) because this transition controls their mechanical properties and allows variation of elastic modulus and viscosity over several orders of magnitude~\cite{Polymer_Fracture}. This is why the question of the polymer dynamics in confined geometries such as thin film gives rise to an important effort from the scientific community. During the last 20 years, researches~\cite{ediger_dynamics_2014} have led to believe that $T_{\mathrm{g}}$ could be lowered or increased by a few tens Kelvin in a 10~nm thick polystyrene (PS) film. This could mean important slowing down or acceleration of the dynamics that would lead to substantial changes in mechanical properties.  However these modifications are still in debate and only partially understood~\cite{ediger_dynamics_2014}. Moreover the influence of the interaction between a supported film and its substrate is still an open question.

The effect of confinement on the $T_{\mathrm{g}}$ of a thin PS film was first reported by Keddie \textit{et al.} in 1994~\cite{keddie_1994}. They measured by ellipsometry a reduction in $T_{\mathrm{g}}$ reaching 25~K for thin PS films of 120, 500 and 2900~kg\(\cdot\)mol$^{-1}$ deposited on the silicon oxide layer covering Si substrates, compared to $T_{\mathrm{g}}$ for the same bulk material. This effect appeared to be independent of the molecular weight and was observed for films thinner than 40~nm. Following this pioneering work, extensive research was conducted on the subject using different experimental techniques (ellipsometry, Brillouin scattering, positron annihilation, {\em etc.}) and for different geometries~\cite{forrest_glass_2001}. It appeared that the reduction in $T_{\mathrm{g}}$ was 
particularly important for free standing films that present two free surfaces. Indeed, this reduction reaches 60~K for 30~nm free standing films. However, unlike supported film, this $T_{\mathrm{g}}$ reduction was found dependent of the molecular weight of the chains~\cite{forrest_1997}. Therefore its origin could be different from that observed by Keddie \textit{et al.}  It has also been verified by Baumchen \textit{et al.}~\cite{baumchen_2012} that when the exact same PS film is transferred from a free standing situation to a supported one, the measured value of $T_{\mathrm{g}}$ is the one usually measured for supported films, indicating that the observed effect does not originate from sample preparation artifacts.

The fact that the reduction of $T_{\mathrm{g}}$ depends on the number of free surfaces in the films suggests that the mobility of the chains near interfaces is an important factor for  understanding  this phenomenon. Indeed, some experiments~\cite{ellison_torkelson_2003, fakhraai_2008, yang_2010, yoon_2014, chai_2014} implied that the value of $T_{\mathrm{g}}$ measured in thin films does not correspond to a sharp transition between a situation where the film is liquid to a situation where the whole film is glassy. The idea has been developed that below the bulk $T_{\mathrm{g}}$,  a liquid-like layer exists at the free interface of the film whose thickness and viscosity depend on the temperature. This idea would be compatible with the observed thickness dependence of the apparent $T_{\mathrm{g}}$~\cite{forrest_2014}.  In 2014, Chai \textit{et al.}~\cite{chai_2014} were able to quantitatively probe the surface mobility of thin films at different temperatures by measuring the shape of a stepped polymer film. They showed that below the bulk $T_{\mathrm{g}}$(bulk) the film is well described by a model where the film is glassy but with a thin liquid layer remaining at the surface.

Ellison and Torkelson~\cite{ellison_torkelson_2003} studied the fluorescence of multilayer films composed of a 10-15~nm thick layer of labeled PS and layers of unlabeled PS. By varying the depth of the labeled layer, they were able to probe the distribution of the local  $T_{\mathrm{g}}$s across the thickness of PS films. They reported a reduction in $T_{\mathrm{g}}$ of 30~K at the free surface of a thick film while they measured the bulk $T_{\mathrm{g}}$ near the substrate and at the middle of the film. For films thicker than 50~nm, this effect was independent of the total thickness of the film. For thinner films  a smaller reduction in the surface $T_g$ was observed.   For a 24~nm film, they measured the same value of $T_{\mathrm{g}}$ near the substrate and near the interface, leading to the conclusion that the free surface mobility is influenced by the presence of the substrate. Indeed, some studies have reported an effect of the film-substrate interaction on $T_g$. Fryer \textit{et al}.~\cite{fryer_2001} measured $T_{\mathrm{g}}$ of thin PS films deposited on self-assembled films of octadecyltrichlorosilane whose interaction with PS could be tuned with an exposure  to X-rays. In the case of high interfacial energy, they reported a rise in the overall $T_{\mathrm{g}}$ when the film thickness was decreased. They showed that for 22~nm thick films, $T_{\mathrm{g}}$ scales linearly with the interfacial energy. More recently, Roth \textit{et al.}~\cite{roth_eliminating_2007} measured $T_{\mathrm{g}}$  of 14~nm thick PS films deposited on thick films of incompatible polymers (PMMA and P2VP)  and found values very close to bulk $T_{\mathrm{g}}$, which is quite different from what is observed for a films deposited on a silicon wafer. 

Some models have been developed in order to explain the observed $T_{\mathrm{g}}$ shifts in polymer thin films. Long and Lequeux in 2001~\cite{long_lequeux_2001} proposed that the glass transition is controlled by the percolation of domains of slow dynamics, which can explain the decrease in $T_{\mathrm{g}}$ due to the finite thickness of a film as well as an influence of a strong interaction at interfaces. Simpler phenomenological models combining 2 or 3 layers (a low $T_{\mathrm{g}}$ layer at the free surface where the mobility is high, a bulk $T_{\mathrm{g}}$ layer in the middle of the film and occasionally a high $T_{\mathrm{g}}$ layer at the substrate interface where the mobility is reduced) have been used to fit the data~\cite{forrest_2000}. More recently, a model developed by Salez \textit{et al}.~\cite{salez_2015} showed that the fact that less cooperation is needed at a free interface for the monomers to move is sufficient to explain the  $T_{\mathrm{g}}$ reductions observed in PS thin films.

The effect of interfacial structure, and in particular the effect of grafted polymer chains, on the thin-film $T_{\mathrm{g}}$, has also been investigated. There are two equivalent ways of characterizing the grafting density of polymer grafted layers. One is the thickness of the dense layer made only of the grafted layer chains (that we call a dry grafted layer), which we refer to as $z^{\ast}$. The second is $\sigma$, the areal density of grafted layer chains, which is obtained by dividing  $z^{\ast}$ by the molecular volume, $Nv_{0}$, where $N$ is the degree of polymerization of the grafted chains and $v_{0}$ is the volume of a repeating unit in the polymer:
\begin{equation}
\sigma=\frac{z^{\ast}}{Nv_{0}}
\end{equation}
Two corresponding dimensionless representations of the coverage can also be defined.  The first of these normalizes $z^{\ast}$  by a characteristic dimension of the entire polymer molecule, often the root-mean-squared end-to-end distance, $R_{0}$ for a Gaussian chain of length $N$ in the melt state. Note that $R_{0}=N^{1/2}a$, where $a$ is the statistical segment length of a monomer. Another possibility is to multiply $\sigma$ by the square of a characteristic monomer size, thereby defining a dimensionless grafting density, $\Sigma$:
\begin{equation}
\Sigma=a^{2}\sigma=\frac{z^{\ast}}{Na} \label{Sigma}
\end{equation}
where we have defined the monomer size $a$ so that $v_{0}=a^{3}$. Note that $\Sigma$ can be viewed as the fraction of potential grafting sites that are occupied, and is always much lower than one for high molecular weight chains~\cite{de_gennes_conformations_1980}. 
When $z^{\ast}$ is lower than $a$, the grafted chains are physicaly separated and are in the so-called mushroom regime. Otherwise, the grafted chains can be unstreched if, while overlaping, they are not close enough to repel each other, or they can be streched if $z^{\ast} > R_{0}$. This limit corresponds to a grafting density $\Sigma_{\mathrm{SL}}$:
\begin{equation}
\Sigma_{\mathrm{SL}}=\frac{1}{N^{1/2}} \label{Sigma_max}
\end{equation}
In grafting-to techniques, in which pre-polymerized end-functionalized chains are attached to the surface, there are two regimes in the kinetics~\cite{ligoure_1990}. First, a fast regime where the grafting density can reach $\Sigma_{\mathrm{SL}}$ and then a slow regime where, in order to graft, the chains have to pass the activation barrier due to the presence of the streched grafted layer.
It appears that in practice~\cite{Auroy_1991, shull_end-adsorbed_1996}, we are limited to the first fast regime. 
Values of $\Sigma$ in excess of $\Sigma_{\mathrm{SL}}$ can be obtained more easily using grafting-from techniques. In this paper we use repeating units with a molecular weight of 104~g.mol$^{-1}$ to define $N$, and use $a=0.67$~nm~\cite{Cotton_1974} to determine $\Sigma$ and $\Sigma_{\mathrm{SL}}$. In our notation the grafted chains have a degree of polymerization of $N$ and a molecular weight of $M_\mathrm{w}^N$, and are in contact with a melt of ungrafted chains with a degree of polymerization of $P$ and a molecular weight of $M_\mathrm{w}^P$. We express the overall grafting density as $\Sigma/\Sigma_{\mathrm{SL}}$, or equivalently, as $z^{\ast}/R_{\mathrm{0}}$.

The effect of grafting on $T_{\mathrm g}$ of thin PS films was first studied by Keddie \textit{et al.} in 1995~\cite{keddie_1995} who used ellipsometry to measure $T_{\mathrm g}$ of naked (no bulk ungrafted chains) PS grafted layers with $M^{\mathrm N}_{\mathrm w}$ = 225~kg\(\cdot\)mol$^{-1}$ and $\Sigma/\Sigma_{\mathrm{SL}} = 0.17 - 0.32$ on Si substrates with a native oxide layer. They reported values of Tg between 335~K and 360~K for 5-7~nm thick layers which can be compared to the 335-340~K that they measured for 9-10~nm thick PS films. By comparing these results to the 325~K that Raegen et al.~\cite{raegen_2008} were able to measure on 6~nm thick PS films, it is possible to conclude in an effect of grafting on $T_\mathrm{g}$ for very thin grafted layers. In 2001, Tsui \textit{et al.}~\cite{tsui_2001} measured  $T_{\mathrm g}$ of 33 nm thick PS films deposited on grafted layers made of random copolymers of styrene and methyl methacrylate of variable styrene fraction ($M^{\mathrm N}_{\mathrm w}$ = 10~kg\(\cdot\)mol$^{-1}$,  $\Sigma/ \Sigma_{\mathrm{SL}} = 0.46$). In the case of a PS film on a grafted layer only made of styrene monomers ($M^{\mathrm P}_{\mathrm w}$ = 96~kg\(\cdot\)mol$^{-1}$), they reported $T_{\mathrm g} = 95 \pm 5$~K, equivalent to what is measured for a film of same thickness deposited on SiO$_2$. 
Tate \textit{et al.}~\cite{tate_2001} studied PS films on grafted PS grafted layers with a high grafting density ($\Sigma/\Sigma_{\mathrm{SL}} = 0.91$, $M^{\mathrm N}_{\mathrm w} = M^{\mathrm P}_{\mathrm w}$ = 100~kg\(\cdot\)mol$^{-1}$). They reported an increase in $T_{\mathrm{g}}$ for thicknesses below 100~nm. This increase reached 25~K for 40~nm films.
In 2010, Lee \textit{et al.}~\cite{lee_2010} studied $T_{\mathrm{g}}$ of PS films ($M^{\mathrm P}_{\mathrm w}$ = 102~kg\(\cdot\)mol$^{-1}$) deposited on grafted layers (3.7-38~kg\(\cdot\)mol$^{-1}$) with varying grafting density ($\Sigma/\Sigma_{\mathrm{SL}} = 0.94-1.31 $). If their data are plotted as a function of the total thickness of the PS film (including the grafted layer), no effect of the molecular parameters of the grafted chains can be seen. However, their measured values of $T_{\mathrm{g}}$ are a few tens of Kelvins below the data from the literature concerning the case of PS films deposited directly on a bare Si wafer. In 2011, Clough \textit{et al}.~\cite{clough_2011} measured $T_{\mathrm{g}}$ of PS films ($M^{\mathrm P}_{\mathrm n}$ = 41k) deposited on PS grafted layers ($M^{\mathrm N}_{\mathrm n}$ = 5, 10 and 96~kg\(\cdot\)mol$^{-1}$, $\Sigma/\Sigma_{\mathrm{SL}}=0.69-0.86$) or on SiO$_{x}$. When plotted as a function of the total thickness of the PS film, again, no evidence of an influence of the grafted chains on $T_{\mathrm{g}}$can be evidenced.
In 2013, Dinelli \textit{et al.}~\cite{dinelli_2013} studied the penetration of an AFM tip at the surface of a film as a function of temperature. From these measurements they determined $T_{\mathrm g}$ of 30~nm  PS films deposited on different substrates (Si-H, SiO$_{x}$) or on grafted layers of variable lengths and grafting density. They reported the same value of $T_{\mathrm g}$ for a film ($M^{\mathrm P}_{\mathrm w}$ = 13~kg\(\cdot\)mol$^{-1}$) deposited on a SiO$_{x}$, a Si-H substrate or on PS grafted layers ($M^{\mathrm N}_{\mathrm w} = 135$~kg\(\cdot\)mol$^{-1}$, $\Sigma/\Sigma_{\mathrm{SL}} = 0.12$). In the case of a film made of long chains ($M^{\mathrm P}_{\mathrm w}$ = 483~kg\(\cdot\)mol$^{-1}$), they reported an influence of the presence of a grafted layer on $T_{\mathrm g}$, from -5~K for a short grafted layer ($M^{\mathrm N}_{\mathrm w} = 7.5$~kg\(\cdot\)mol$^{-1}$, $\Sigma/\Sigma_{\mathrm{SL}} = 0.53$) to +20~K for a long grafted layer ($M^{\mathrm N}_{\mathrm w} = 135$~kg\(\cdot\)mol$^{-1}$, $\Sigma/\Sigma_{\mathrm{SL}} = 0.12$). 
Finally, in 2015 Lan \textit{et al.}~\cite{lan_2015} used ellipsomery to measure $T_{\mathrm g}$ of dense PS naked grafted layers, called brushes, made via a grafting from technique. They made different brushes from $M^{\mathrm N}_{\mathrm n}$ = 23~kg\(\cdot\)mol$^{-1}$ to $M^{\mathrm N}_{\mathrm n}$ = 170~kg\(\cdot\)mol$^{-1}$, $\Sigma/\Sigma_{\mathrm{SL}}>1 $. They reported no effect of the confinement on the average $T_{\mathrm g}$ for brushes thicker than 11~nm, in contrast to the  $T_{\mathrm g}$ reduction that is observed for non-grafted PS films in this thickness range. They also measured the spatial distribution of $T_{\mathrm g}$s across the thickness of the films. They found a strong heterogeneity with values 20~K lower than the bulk at the free surface and 36~K higher than the bulk near the substrate. This has to be compared to the bulk $T_{\mathrm g}$ that was measured near the substrate in the case of non-grafted films~\cite{ellison_torkelson_2003}.

\begin{figure}[htbp]
  \centering
  \includegraphics[width=220pt]{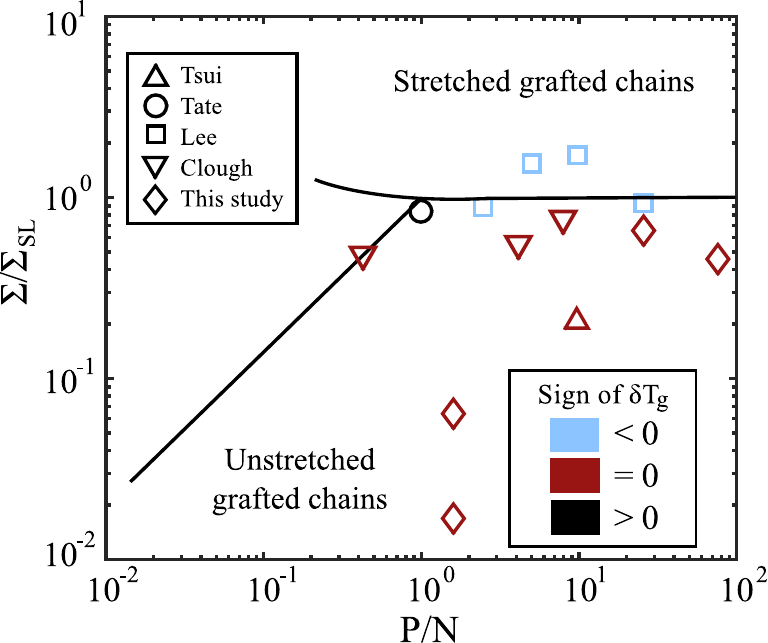}
  \caption{Diagram summarizing the expected conformations and interpenetration between grafted chains (with degree of polymerization $N$) and bulk chains (with degree of polymerization $P$) regimes~\cite{de_gennes_conformations_1980}. The locations in this diagram of the samples investigated by Tsui \textit{et al.}~\cite{tsui_2001}, Tate \textit{et al.}~\cite{tate_2001}, Lee \textit{et al.}~\cite{lee_2010} and Clough \textit{et al.}~\cite{clough_2011} are shown. The different colors show the effect of the presence of the grafted layers on  $T_{\mathrm g}$ : black in the case of an augmentation compared to a film having the same total thickness deposited on a bare silicon substrate, red for no effect and blue for a reduction. The red diamonds represent the systems studied in this article.}
  \label{diagramme_biblio}
\end{figure}

To summarize, several authors~\cite{tate_2001,tsui_2001,lee_2010,clough_2011, lan_2015} have reported an influence of the presence of grafted chains on $T_{\mathrm{g}}$ of thin PS films in the case of dense grafted layers ($\Sigma/\Sigma_{\mathrm{SL}} \gtrsim 1 $) in the so-called dry-grafted layer regime ($M^{\mathrm P}_{\mathrm w}>M^{\mathrm N}_{\mathrm w}$), where the matrix chains do not substantially overlap with the grafted layer chains. We define $\delta T_\mathrm{g}$ as the difference between the average $T_\mathrm{g}$ of a grafted layer covered with a film and $T_\mathrm{g}$ of a film with the same total thickness directly deposited on a wafer. A schematic summary of the location of these previous investigated systems in a map diagram showing the boundaries between the different regimes of conformation and interpenetration between grafted and bulk chains, as deduced from de Gennes~\cite{de_gennes_conformations_1980} and where the sign of $\delta T_\mathrm{g}$ appears is shown in Figure~\ref{diagramme_biblio}.

The aim of the present article is to investigate more precisely the glass transition of naked grafted chains and to explore the role of these grafted chains on the average $T_{\mathrm{g}}$ in a regime of lower grafting densities than in the previous studies, so that a potential effect due to the interpenetration between grafted and bulk chains could be detected. We want to understand how the presence of grafted chains, whose mobility is reduced compared to that of free chains~\cite{chenneviere_2013}, does affect the overall $T_{\mathrm{g}}$ of thin films.  By varying experimentally the grafting density and the grafted chains length it is possible to control the grafted layer-film interaction and interpenetration. Numerical calculations using self consistent field theory (SCFT) have been used to systematically estimate the penetration of a grafted layer in a film~\cite{shull_1991, jones_1992}. By measuring $T_{\mathrm{g}}$ of PS films deposited on polymer grafted layers of different molecular parameters, we can compare the situation of a grafted layer fully penetrated by bulk chains to the case of a grafted layer confined near the substrate and interacting with the film only in a small fraction of its total thickness. In this paper, we present $T_{\mathrm{g}}$ measurements of naked PS grafted layers of variable length and grafting density below $\Sigma_{\mathrm{SL}}$  and of PS films deposited and interdigitated with these grafted layers, and we compare these results to the case of PS films directly deposited on a Si substrate.

\section{Experimental section}

\begin{figure}[htbp]
  \centering
  \includegraphics[width=240pt]{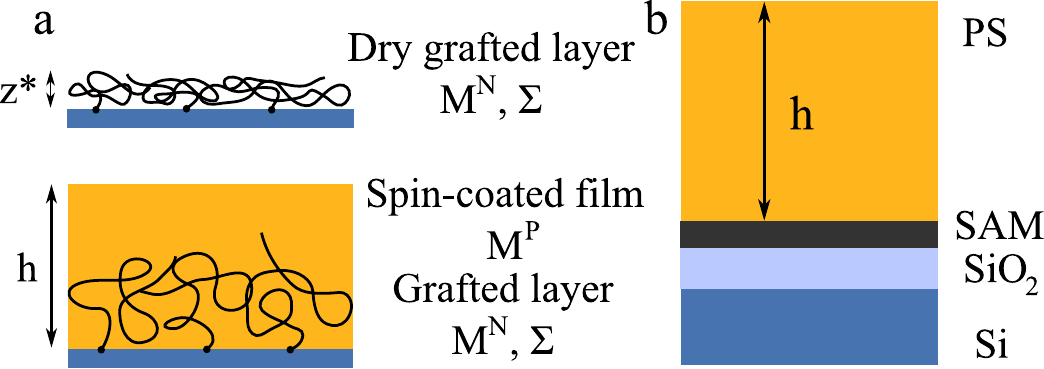}
  \caption{a. Structure of the samples made for this study. $M^{\mathrm{N}}$ is the molecular weight of the grafted chains, $\Sigma$ is the dimensionless grafting density and $M^{\mathrm{P}}$ is the molecular weight of the free chains deposited on the grafted layers. b. Model used for the inversion of the ellipsometric data. It consists of a layer of PS of thickness $h$, which is the free parameter in the inversion, a layer of self assembled monolayer of silane (SAM) and a layer of native silicon oxide whose thicknesses have been measured during the fabrication of the sample, on top of a semi infinite silicon substrate. The corresponding refractive indices at 25~\degres C were obtained from the literature.}
  \label{schema_complet}
\end{figure}
 
\textbf{Grafted layer preparation and characterization.} Figure~\ref{schema_complet} shows the structure of the samples that were made in this study. In order to graft PS chains to the substrate ~\cite{chenneviere_2013}, we used diethoxy (3-glycidyloxypropyl) methylsilane (97\%, Sigma) to form a self-assembled reactive monolayer (SAM) on the native oxide layer of a silicon wafer. The silicon wafers (Si-Mat, Kaufering, Germany) were cleaned by UV ozone treatment~\cite{vig_uvozone_1985} for 45~min. Silicon wafers were put in a desiccator under vacuum containing also 1~mL of silane. The system was then carefully heated ($\geq$ 150~\degres C) to vaporize the silane and to allow the silane deposition and grafting. After 8~h, the wafers were rinsed several times with anhydrous toluene and dried under vacuum. On the top of the SAM monolayer a thin layer of $\alpha$-aminopolystyrene with number-average molecular weight $M_{\mathrm{n}}^{\mathrm{N}}$, was spin-coated at 2000~rpm from a 3~wt \% solution in toluene (polymers with $M_{\mathrm{n}}^{\mathrm{N}}$ = 218~kg\(\cdot\)mol$^{-1}$ and 18~kg\(\cdot\)mol$^{-1}$ were synthesized as descibed previously in~\cite{chenneviere_2013,chenneviere_2016} and a polymer with $M_{\mathrm{n}}^{\mathrm{N}}$ = 5~kg\(\cdot\)mol$^{-1}$ was bought from Polymer Source). The samples were then annealed in a vacuum oven at 140~\degres C to allow the NH$_{2}$ groups to chemically bind to the surface-tethered epoxy groups. The grafting density for each molecular weight of the grafted chains, is efficiently controlled by the annealing time $t_\mathrm{reaction}$ (see Table~\ref{table_brosses}). After rinsing several times with toluene and drying under vacuum to remove the remaining solvent, the thickness $z^{*}$ of the PS grafted layer was measured by ellipsometry. This thickness was found to be between 5 and 12~nm depending on the annealing time and polymer molecular weight $M_{\mathrm{n}}^{\mathrm{N}}$. We recall that, because the grafting-to procedure is involved, $z^{*}$ cannot be higher than $\sim R_0$, which means that $\Sigma$ cannot be larger than $\sim \Sigma_{\mathrm{SL}}$. Table~\ref{table_brosses} summarizes the molecular parameters of the grafted layers prepared for this study. 

Thin layers of PS (5 - 100~nm) of number-average molecular weight $M_{\mathrm{n}}^{\mathrm{P}}$ = 422~kg\(\cdot\)mol$^{-1}$, $M_{\mathrm w}/M_{\mathrm n} = 1.05$ (Toyo Soda Manufacturing, Japan) were then deposited on the grafted layers, by spin-coating a solution of PS in toluene onto a naked grafted layer. The samples were then annealed under vacuum at 140~\degres C for 24~h in order to reach grafted layer/film equilibrium interdigitation~\cite{chenneviere_2013}. The cooling is done by powering off the oven, thus the rate of cooling is not exactly constant but stays between 1 and 0.8~K/min. Films deposited directly on a substrate without grafted chains were also made by spin-coating a solution of PS in toluene on the native oxide layer of a silicon wafer previously cleaned by UV ozone. These samples were then annealed under vacuum  at 140~\degres C for 24~h, similarly to the films deposited on a grafted layer. All samples were stored at ambient temperature under ambient atmosphere and used less than a week after their preparation. In order to quantify the absorption on the substrate, several samples were rinsed  with toluene and the thickness of the remaining layer of PS was measured by ellipsometry and found to have a thickness of 1-2~nm for PS films on silicon oxyde.

\begin{table}[h]
\begin{center}   \begin{tabular}{c| c| c| c| c|c| c}
     Name & $M_{\mathrm n}^{\mathrm N}$ & $M_{\mathrm w}$ & $t_\mathrm{reaction}$ & $z^{*}$ & $\Sigma$ & $\Sigma$ \\
      & (kg\(\cdot\)mol$^{-1}$) & $/M_{\mathrm n}$ & (h) & (nm) & & $/\Sigma_{\mathrm{SL}}$ \\ \hline
     218k-8.8 nm & 218  & 1.28 & 22 & 8.8 & 0.0076 & 0.29 \\
     215k-4.5 nm & 218 &  1.28 & 2.2 & 4.5 & 0.0039 & 0.18 \\
    15k-7.0 nm & 15.5 & 1.12 & 27.5 &7 & 0.085 & 0.86 \\
     5k-3.4 nm & 5 & 1.17 & 24 & 3.4 & 0.13 & 0.73 \\
   \end{tabular}
 \end{center}
\caption{Molecular parameters of the grafted layers made during this study.  $M_{\mathrm{n}}^{\mathrm N}$ is the number-average molar mass of the grafted chains, written in g\(\cdot\)mol$^{-1}$.  $M_{\mathrm w}/M_{\mathrm n}$ is the  chain dispersity. $t_\mathrm{reaction}$ is the time of the grafting reaction. $z^{*}$ is the thickness of the naked grafted layers measured by ellipsometry. $\Sigma$ is the dimensionless grafting density and $\Sigma_{\mathrm{SL}}$ the limit grafting density of the dry grafted layer regime.}
\label{table_brosses} 
\end{table}

To summarize, we produced classical spin-coated films on bare Si wafers, naked grafted layers of low and high molecular weights PS, and thin PS films on short or long PS grafted layers, with various grafting densities.

\textbf{Techniques.} The film thicknesses and glass transition temperatures were measured using a commercial ellipsometer, Accurion EP3, at 658~nm, at an incidence angle of 70\degres. A sample was deposited on a heating stage controlled by a Lakeshore controller. A ramp at 3~\degres C\(\cdot\)min$^{-1}$ was applied from 30~\degres C to 150~\degres C and the ellipsometric angles $\Delta$ and $\Psi$ were monitored during the heating which is done under ambiant air atmosphere. The thicknesses of the films were determined by fitting the ellipsometric data at room temperature using the layer model described in Figure~\ref{schema_complet}. For each sample, the thickness of the oxide layer and of the SAM layer were measured by ellipsometry prior to the deposition of the PS film and are not adjustable parameters in this model. The typical thicknesses were 1.8~nm for the SiO$_ {2}$ layer and 0.5~nm for the SAM. The considered complex refractive indices at room temperature were $n_{\mathrm{Si}} = 3.833+0.014i$~\cite{Green_1995}, $n_{\mathrm{SiO}_{2}} = 1.456$~\cite{Malitson_1965}, $n_{\mathrm{SAM}} = 1.431$ (data from Sigma at 589~nm) and $n_{\mathrm{PS}} = 1.586$~\cite{Sultanova_2009}.

\section{Results and Discussion}

\begin{figure}[htbp]
  \centering
  \includegraphics[width=220pt]{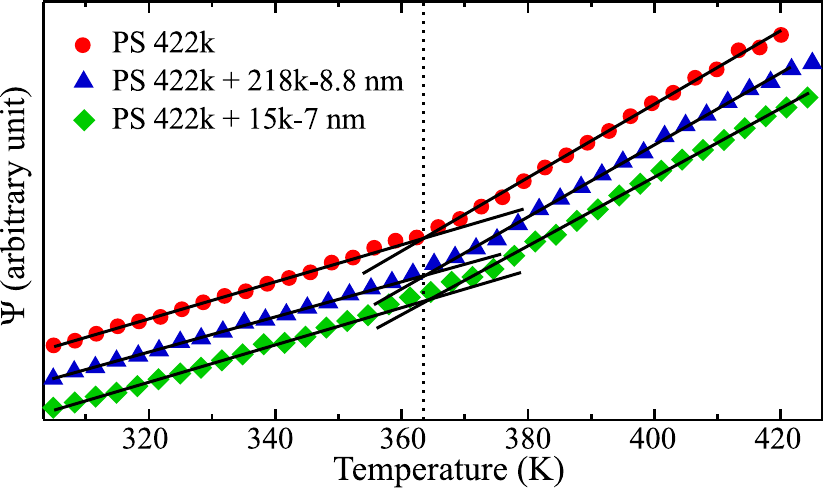}
  \caption{Evolution of $\Psi$ versus the temperature for three PS films having approximately the same thickness, but various film/substrate configurations. The red circles correspond to a 26~nm thick PS film deposited directly on a Si substrate. The blue triangles and the green diamonds correspond to films interdigitated with grafted layers 218k-8.8~nm and 15k-7.0~nm respectively (cf table~\ref{table_brosses}) with total thickness 30~nm and 26~nm. The black lines are the best adjustment of the linear parts. The vertical line shows the value of $T_{\mathrm g}$ defined as the cross over temperature between low and high expansion coefficients.}
  \label{ellipso_films}
\end{figure}

The temperature dependence of the ellipsometric angle $\Psi$ for a 26~nm thick PS film deposited on a Si substrate, is shown in Figure~\ref{ellipso_films} (red circles). This ellipsometric angle depends on the thicknesses and refractive indices of the different layers composing the sample. The thermal expansivity in the melt and in the glassy state being different, the slope of the thickness versus the temperature is different below and above $T_{\mathrm g}$. As these thickness variations are very weak, the problem can be linearized and this modification of slope can be directly measured on the $\Psi(T)$ curve. The observed kink in the $\Psi(T)$ curve is currently admitted to correspond to $T_{\mathrm g}$. As usually done in the literature~\cite{keddie_1994, raegen_2008,lee_2010,clough_2011, baumchen_2012}, we base our measurements on the $\Psi(T)$ curve, less sensitive to noise than $\Delta(T)$. The contrast of the kink, that is to say the relative difference of the slopes below and above $T_{\mathrm g}$ decreases with the film thickness~\cite{forrest_glass_2001}. This is why it is not possible to measure $T_{\mathrm g}$ for PS film thinner than 10~nm by this approach, due to the noise. For the data shown in Figure~\ref{ellipso_films}, the measured value of $T_{\mathrm g}$ is $364\pm1.5$~K, reduced in comparison to the value 373~K that was measured for a 120~nm thick film, and which is also the value reported in the litterature for the bulk $T_{\mathrm g}$~\cite{handbook_polymers}.

\begin{figure}[h]
  \centering
  \includegraphics[width=240pt]{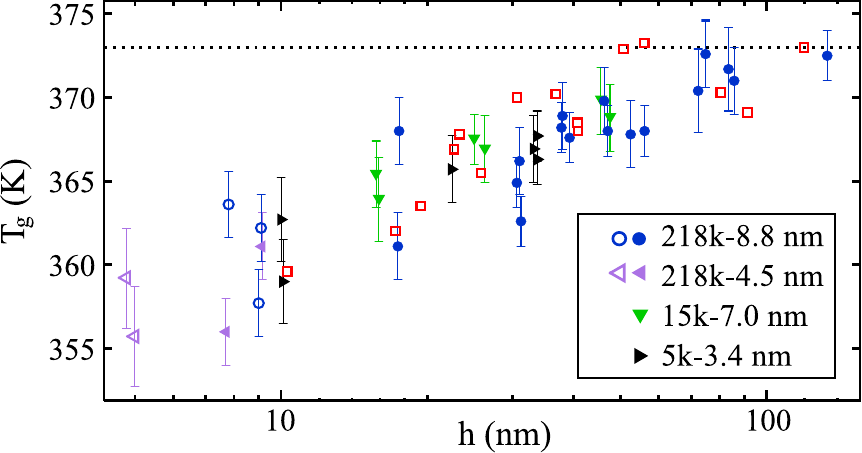}
  \caption{Glass transition temperature measured by ellipsometry for PS films deposited on Si subtrate (red squares), PS films deposited on grafted layers (solid symbols) and naked grafted layers (empty markers). Grafted layers 218k-8.8~nm are in blue, 218k-4.5~nm in violet, 15k-7.0~nm in green and 5k-3.4~nm in black (cf table~\ref{table_brosses}). The horizontal dashed line is the bulk $T_{\mathrm g}$.}
  \label{Tg_brosses}
\end{figure}

Figure~\ref{Tg_brosses} shows all $T_{\mathrm g}$s of the different samples configurations measured in this study, as a function of the overall film thickness. The PS films deposited directly on a substrate without grafting (red squares), for total thicknesses in the range 100~nm to 10~nm present a decrease in the apparent $T_{\mathrm g}$ with the decrease in the film thickness. These data are in quantitative agreement with data from the literature~\cite{forrest_glass_2001, raegen_2008}, which validates our experimental protocol. 

{\bf Naked grafted layers.} The blue and violet open symbols represent $T_{\mathrm g}$ measurements for naked grafted layers 218k-8.8 nm and 218k-4.5 nm. Surprisingly, we could detect a kink on these data despite the fact that the corresponding films are very thin (5 - 9~nm). We do not have a clear interpretation for this last observation. Nevertheless, the main result is that for these naked grafted layers, we did observe a decrease in $T_{\mathrm g}$ consistent with the similar naked grafted layers studied by Keddie~{\itshape et al.}~\cite{keddie_1995}. In the case of 10~nm thick films, we can directly compare this to our measurements for films directly deposited on the substrate and we cannot conclude in any effect of grafting. For thinner grafted layers, it is worth noticing that we report $T_{\mathrm g}$ values higher than the value of $325$~K which Raegen~{\itshape et al.}~\cite{raegen_2008} measured in the case of non-grafted 6~nm thick films.

\begin{figure}[htbp]
  \centering
    \includegraphics[width=240pt]{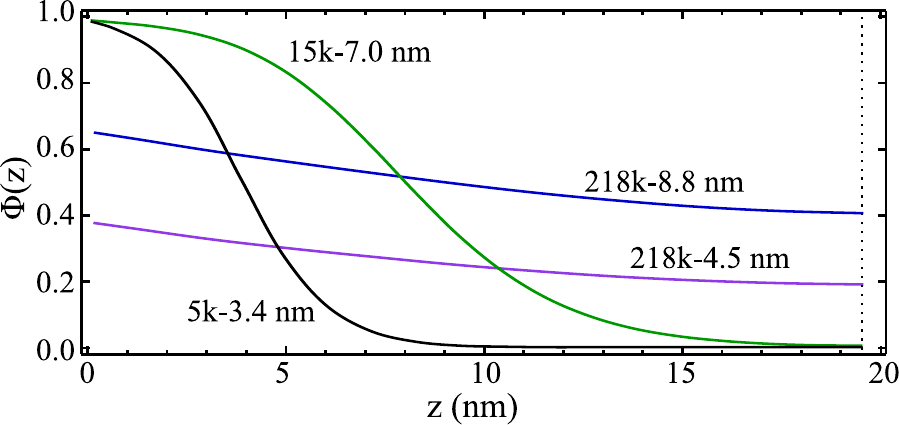}
  \caption{Density profiles calculated by SCFT of grafted layers in contact with a melt of molecular weight $M^{\mathrm{P}}_{\mathrm n}$ = 422~kg\(\cdot\)mol$^{-1}$. The total thickness of the film is set at 19.5~nm, $z$ is the distance from the substrate. These profiles correspond to the grafted layers described in Table~\ref{table_brosses}.}
  \label{SCF_brosses}
\end{figure}

{\bf Role of the interdigitation between grafted and free chains.} We wanted to trace back the role of the coupling between the slow dynamics of the end-tethered chains and the faster dynamics of the free chains in different interdigitation configurations on the $T_{\mathrm g}$ of the films. Indeed, if the large scale motions of a chain in an entangled polymer are correctly described by the reptation model with a long relaxation time given by the reptation time, the dynamics of a surface attached chain is different and the long term motions are closer to the retraction time of an arm in a star polymer. A recent study of the interdigitation dynamics showed that the healing time of an interface is dominated by the exponentially long retraction time of the end-attached chains mediated by the reptation of the melt~\cite{chenneviere_2013}. In order to measure the effect of the couplings between the slow dynamics of the end-attached chains and the chains deposited on top of the grafted layer, we chose samples with varying surface grafting densities, in front of the same melt, and at fixed thickness. More specifically, grafted layers named 218k-8.8~nm and 218k-4.5~nm are made of long chains and are not densely grafted. The interpenetration between these grafted layers and the film is then very important. grafted layers 15k-7.0~nm and 5k-3.4~nm are more densely grafted and made of short chains. In this case, the entropic cost of stretching of the grafted chains associated to the penetration of the melt inside the grafted layer is high~\cite{de_gennes_conformations_1980}, leading to weak interpenetration. The equilibrium interpenetration between a grafted layer and a melt can be quantified by calculating the density profile $\Phi$ of monomers belonging to the end-grafted chains as a function of the distance to the wall $z$, for different grafting densities, molecular weights of the chains and  resulting overall thickness $h$ of the film. We used a previously described program to compute these density profiles using self consistent field theory~\cite{shull_1991}. The size of a monomer used in this computation was $a$ = 0.67~nm. Figure~\ref{SCF_brosses} shows these density profiles for the grafted layers used in this study in contact with the same melt forming a 19.5~nm thick film.

In Figure~\ref{ellipso_films}, we present the raw ellipsometric data $\Psi$ versus the temperature in the case of different films of overall thicknesses close to 30 nm. The blue triangles and the green diamonds represent $\Psi(T)$ for films interdigitated with grafted layers 218k-8.8 nm and 15k-7.0 nm. The total thicknesses (30~nm and 26~nm) are close to that of the film made of free chains (red circles). The position of the kink is the same for the three films, with a decrease of the apparent $T_{\mathrm g}$ compared to bulk, but {\em no noticeable influence of the presence of grafted chains in the films}.

More generally it appears on our results (Figure \ref{Tg_brosses}) that $T_{\mathrm g}$ does not seem to depend on the molecular characteristics of the grafted layer on which the film is deposited. Moreover, \textit{we do not see any difference between the films deposited directly on a bare Si substrate and the films deposited on a grafted layer while, of course, we do observe a dependence of $T_{\mathrm g}$ upon the overall film thickness}. 

\begin{figure}[h]
  \centering
  \includegraphics[width=220pt]{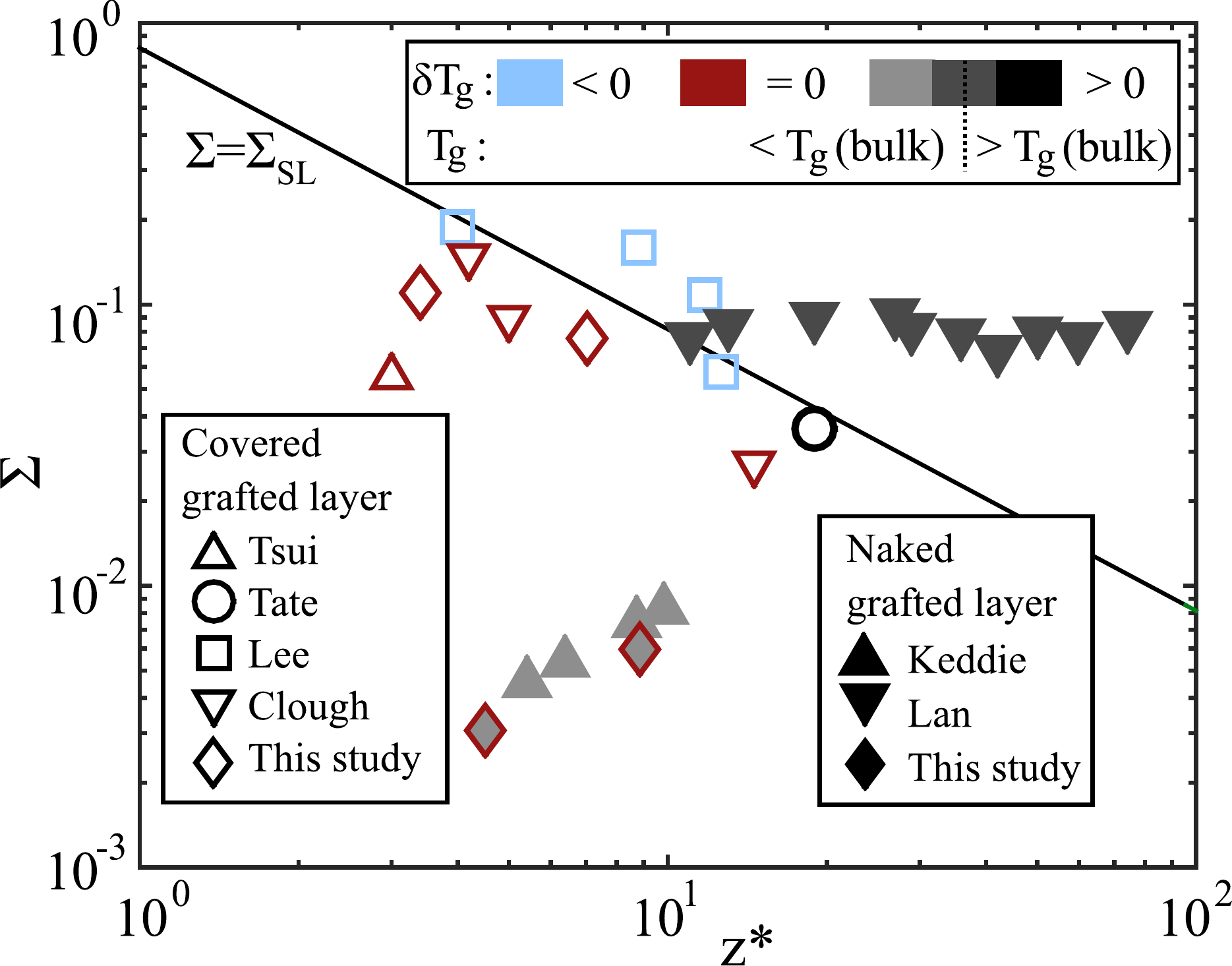}
  \caption{Dimensionless grafting density $\Sigma$ versus the thickness of the dry grafted layer $z^{*}$ for all data in the literature: data from Keddie \textit{et al.}~\cite{keddie_1995}, Tsui \textit{et al.}~\cite{tsui_2001}, Tate \textit{et al.}~\cite{tate_2001}, Lee \textit{et al.}~\cite{lee_2010}, Clough \textit{et al.}~\cite{clough_2011} and Lan \textit{et al.}~\cite{lan_2015} are shown. Solid symbols correspond to naked grafted layers. The green line represents the practical limit of $\Sigma$ for a grafting-from approach, with $z^{*}=z_{\mathrm{SL}}^*$ as given by equation ~\ref{z max star}. As in Figure~\ref{diagramme_biblio}, the different colors show the effect of the presence of the grafted layers on  $T_{\mathrm g}$, with blue symbols for a decrease in $T_{\mathrm g}$ due to the grafted chains, red, no effect on $T_{\mathrm g}$, and grey to black, an increase in $T_{\mathrm g}$. Light grey corresponds to the case where a reduction in $T_{\mathrm g}$ compared to the bulk is still observed while black corresponds to an augmentation.  The diamonds represent the samples studied in this article.}
  \label{diagramme_biblio_KS}
\end{figure}

{\bf Comparison with the literature.} When interpreting these results it is important to keep in mind the fact that the glass transition reflects changes in the dynamics at local scales. These segmental dynamics are {\em a priori} coupled to the local conformations of both the free and grafted chains. It is expected that the local conformation of the grafted chains should be affected by the grafting density, especially for grafting densities larger than $\Sigma_{\mathrm{SL}}$ where they are forced to stretch in the direction normal to the substrate. This stretching, if strong enough, could affect the segmental dynamics and the $T_\mathrm{g}$ quite differently from what happens for grafted chains below $\Sigma_{\mathrm{SL}}$. Indeed, in this last case, the grafted chains tend to keep their Gaussian configuration when equilibrated with a melt or to flatten down to the surface for a dry grafted layer. In order to examine if such trends were visible in all data reported in the literature, we have redrawn the Figure~\ref{diagramme_biblio}, in a graph where the data are placed depending on the grafting density as a function of the dry thickness of the grafted layer, $z^{*}$. Contrary to Figure~\ref{diagramme_biblio}, the data corresponding to naked grafted layers can now appear in this figure. The black line locates the upper limit of grafting-to techniques, with $\Sigma=\Sigma_{\mathrm{SL}}$ as a function of $z_{\mathrm{SL}}^{*}$ obtained by equating $\Sigma$ to $\Sigma_{\mathrm{SL}}$ and eliminating $N$ from equations~\ref{Sigma} and~\ref{Sigma_max}:
\begin{equation}
\Sigma=\frac{a}{z_{\mathrm{SL}}^{*}} \label{z max star}
\end{equation}
It appears clearly in Figure~\ref{diagramme_biblio_KS} that the data from Lan \textit{et al.} correspond to a quite different grafting regime than all other data, well above $\Sigma_{\mathrm{SL}}$. Then the perturbations to the chain dynamics associated with the grafting are strong enough to overcome the antagonist effect of the free surface (which is expected to accelerate the dynamics, due to the corresponding local loss of constraints on the monomer motions). This explanation is consistent with the spatial distribution of $T_{\mathrm g}$s that they measured in naked grafted layers compared to non-grafted films. It is also consistent with the reduction of $T_{\mathrm g}$ that they reported for a surface layer deposited on a brush when the bush thickness is increased. It illustrates that the impact of grafting can propagate some tens of nanometers away from the substrate and measurably impact the average $T_{\mathrm g}$ in case of high enough grafting densities. This could help us to understand why an effect of grafting seems to appear only in very thin grafted layer in the range of lower grafting densities that we used in this study.
Studies from Tsui {\itshape et al.}, Clough {\itshape et al.}, consistent with this one, show no differences in $T_{\mathrm g}$ between films deposited directly on the substrate of films deposited on a grafted grafted layer. It does not seem possible however to rationalize all the literature using these arguments. Indeed the work done by Tate {\itshape et al.} shows a very different behavior although $\Sigma/ \Sigma_{\mathrm{SL}}$ is not much higher than in the other studies. On the other side, data from Lee {\itshape et al.} on films deposited on grafted layers of rather short molecular weights, and relatively densely grafted show a decrease in $T_{\mathrm g}$, in a situation where only a quite weak interpenetration between free and grafted chains is expected compared to the previously cited studies. It may well be that in this case, the short grafted chains screen down the interactions with the substrate and leads to an overall faster dynamics.

Indeed, a second important fact to keep in mind when trying to rationalize these results is the fact that, in the case of a grafted layer of low grafting density, chains from the melt can penetrate into the grafted layer and interact with the substrate. If an attractive interaction does exist between the polymer chains and the substrate, then an adsorbed layer forms at the substrate / film interface. This could explain why we do not observe any differences in $T_{\mathrm g}$ between films directly deposited on the substrate and films deposited on grafted layers. Some indication that such an adsorbed layer is indeed present in our samples, where PS is directly deposited on the silicon substrate, is the observation of a residual 1-2~nm thick PS layer after rinsing thoroughly the substrate.
Because the dynamics involved in $T_{\mathrm g}$ correspond to local motions of the monomers, the perturbations associated to the presence of an interface act over a limited range. The effect of the conformational changes associated with the grafting on the glass transition appears small, and generally overwhelmed by substrate and surface effects that affect both grafted and ungrafted chains in a similar way, except for large grafting densities leading to strongly stretched grafted chains, i.e. when $\Sigma$ and $z^{*}$ are both large, as for the data by Lan shown in Figure~\ref{diagramme_biblio_KS}.

Finally, a quite puzzling result that we are not able to interpret right now is the fact that we were able to still measure a $T_{\mathrm g}$ through ellipsometric techniques on this very dry grafted layers, well below the 10~nm thickness usually  considered as a limit for such measurements. It would be interesting to check directly the exact conformation of the chains involved in such grafted layers, for example through neutrons reflectivity, in order to try to correlate the observed $T_{\mathrm g}$ to the chains conformation.

\section{Conclusion}

We measured by ellipsometry the apparent $T_{\mathrm g}$ of PS films supported on the native oxide layer of a Si substrate, naked PS grafted layers of different grafting densities and chain lengths, and grafted layers with non-grafted PS overlayers. Using SCFT, we calculated the density profile of the grafted layers in contact with a melt and we verified that our samples showed two different regimes: a regime where the grafted chains fully penetrate the film and are present at the free interface and a regime close to the dry grafted layer limit where the grafted chains are confined near the substrate. In agreement with the literature, we find that for PS films deposited on a Si substrate, $T_{\mathrm g}$ decreases when the thickness of the film is reduced. This diminution is observable for films thinner than 40~nm and reaches 13~K for a thick 10~nm film. Measured values of $T_{\mathrm g}$ did not show any dependence on the molecular parameters of the grafted layers on which the films were deposited. Moreover, we did not see any difference between films deposited on grafted layers or on a Si substrate. On the other side, for very thin grafted layers the magnitude of the reduction in $T_{\mathrm g}$ that we measured is not as high as what is reported in the literature for a PS film of same thickness~\cite{raegen_2008}. This could be an evidence of an effect of grafting on $T_{\mathrm g}$ appearing for very thin layers. Our overall conclusion is that if grafting can lead to an increase in $T_{\mathrm g}$, grafted layers obtained by a grafting-to approach, where pre-polymerized chains are grafted to the surface, fall within the parameter space where the glass transition is dominated by free surface effects, so that no neat increase is observe in average $T_{\mathrm g}$.

\section{Acknowledgments}
We thanks the ERC starting grant (agreement 307280-POMCAPS) program and the ANR-ENCORE program (ANR-15-CE06-005) for funding this research.  KRS acknowledges support from the DMR Polymers program of the National Science Foundation from grant DMR-1410968.

\bibliography{biblio}

\newpage
Graphical abstract :

\centering
\def\svgwidth{200pt}
\includegraphics{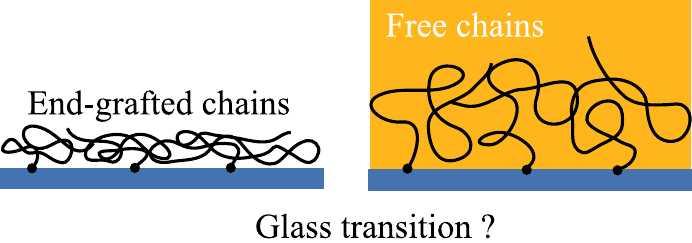}

\end{document}